\documentclass[aps,twocolumn,prl,superscriptaddress,showpacs]{revtex4}

\usepackage{amsmath,amssymb,bm}
\usepackage{graphicx}
\usepackage{epstopdf}
\usepackage{latexsym}
\usepackage{subfigure}
\usepackage{color}

\newcommand{\rvec}{\mathbf{r}}
\newcommand{\qvec}{\mathbf{q}}
\newcommand{\hc}{\mathrm{h.c.}}

\begin{document}

\title{Exotic Gapless Mott Insulators of Bosons on Multi-Leg Ladders}
\author{Matthew S. Block}
\author{Ryan V. Mishmash}
\affiliation{Department of Physics, University of California, Santa Barbara, California 93106, USA}
\author{Ribhu K. Kaul}
\affiliation{Department of Physics and Astronomy, University of Kentucky, Lexington, Kentucky 40506, USA} 
\author{D. N. Sheng}
\affiliation{Department of Physics and Astronomy, California State University, Northridge, California 91330, USA}
\author{\\ Olexei I. Motrunich}
\affiliation{Department of Physics, California Institute of Technology, Pasadena, California 91125, USA} 
\author{Matthew P. A. Fisher}
\affiliation{Department of Physics, University of California, Santa Barbara, California 93106, USA}
\affiliation{Department of Physics, California Institute of Technology, Pasadena, California 91125, USA} 

\date{\today}

\begin{abstract}
We present evidence for an exotic gapless insulating phase of hard-core bosons on multi-leg ladders with a density commensurate with the number of legs.  In particular, we study in detail a model of bosons moving with direct hopping and frustrating ring exchange on a 3-leg ladder at $\nu=1/3$ filling.
For sufficiently large ring exchange, the system is insulating along the ladder but has two gapless modes and power law transverse density correlations at incommensurate wave vectors.  We propose a determinantal wave function for this phase and find excellent comparison between variational Monte Carlo and density matrix renormalization group calculations on the model Hamiltonian, thus providing strong evidence for the existence of this exotic phase.  Finally, we discuss extensions of our results to other $N$-leg systems and to $N$-layer two-dimensional structures.
\end{abstract}


\maketitle

Creating a complete catalogue of the possible zero-temperature phases of interacting bosons in low-dimensional lattice systems is a fundamental open problem in condensed matter physics~\cite{wen}. The last few decades have seen a number of proposals for new entries in this catalogue: the gamut of phases now runs from the well-studied ``solids'' and ``superfluids'' to exotic possibilities such as variants of a ``Bose metal''~\cite{feigelman,phillips,dbl_0,dbl_2leg}.  Equally challenging is the identification of parent Hamiltonians that realize each of the entries as a ground state. This understanding will clearly aid in the search for novel many-body physics in, for example, synthesized materials~\cite{leon} and ultracold atomic gases in optical lattices~\cite{optlatt}.

\begin{figure}[b]
\vspace{-0.1in}
\includegraphics[scale=0.26]{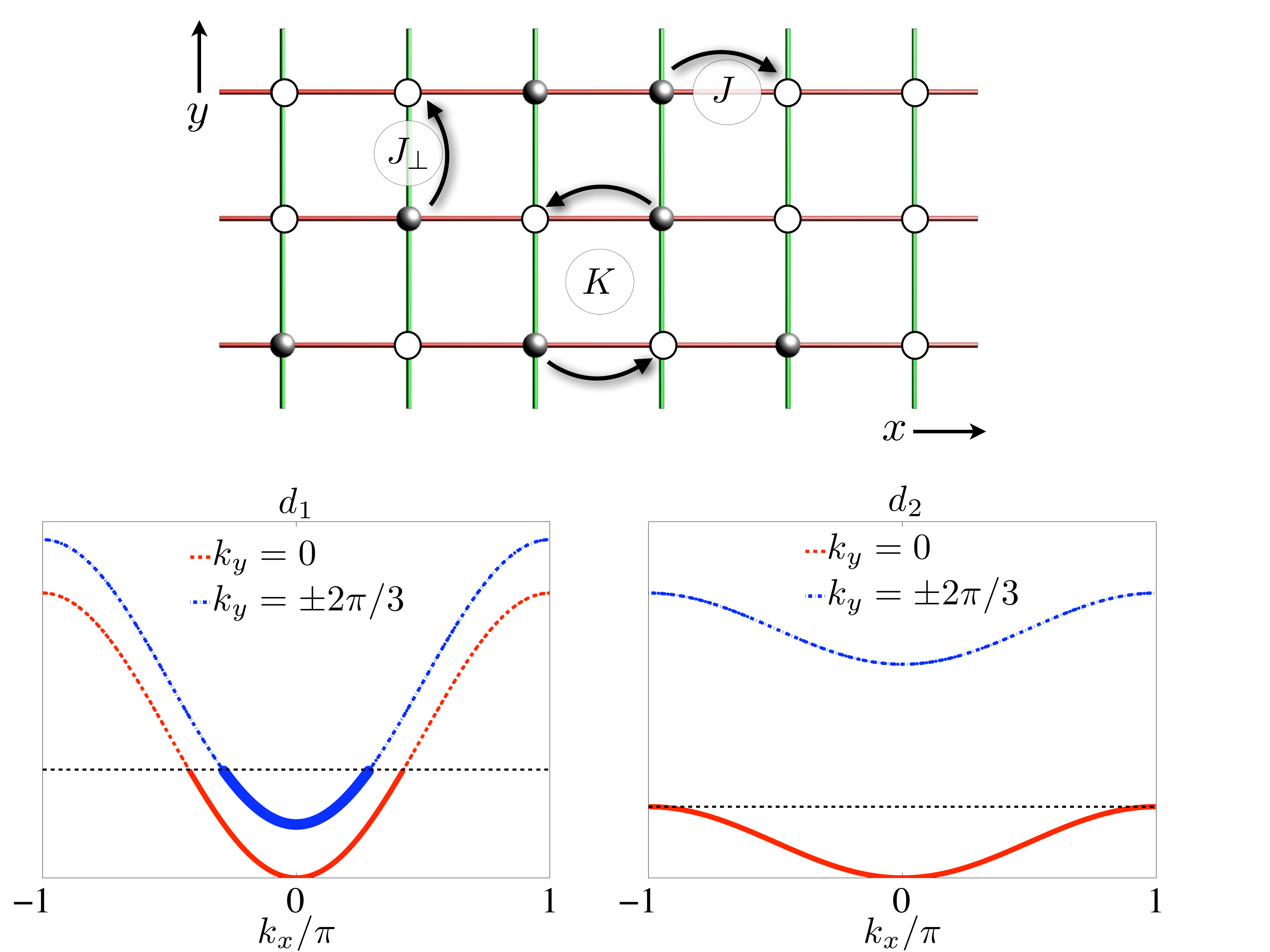}
\caption{(color online) Top: Illustration of the 3-leg ladder geometry with periodic boundary conditions in both the $x$ and $y$ directions.  The hopping strength is $J$ in the $x$ direction and $J_\perp$ in the $y$ direction; the ring term $K$ moves a pair of hard-core bosons on diagonal corners of a plaquette to the other diagonal corners (if empty). Bottom: The $d_1$ and $d_2$ bands for the 3-leg system at $\nu=1/3$ are shown for the gapless Mott insulator studied here. There are exactly enough bosons so that the $d_2$ fermions fill up one band completely.  The bold curves indicate occupied momentum states, and note that there are two degenerate bands at $k_y=\pm 2\pi/3$.  \label{fig:3ll}}
\end{figure}

Weakly interacting bosons generally condense forming a superfluid at low temperatures.  However, for strong interactions at noninteger commensurate densities, bosons naturally break translational symmetries of the lattice, crystallizing into a solid. It is also possible for these two orders to coexist, resulting in the formation of a ``supersolid'' for the bosons.  At integer densities, there is also the possibility of a featureless ``Mott insulator'' that has a nondegenerate ground state with a gap to all excitations.  These phases have been demonstrated to exist convincingly in numerical simulations of lattice models~\cite{sandvik,triss,krauth}.  States that break time reversal symmetry (``chiral spin liquid''~\cite{laugh}) or lattice rotational symmetry (``nematic''~\cite{read}) have also been proposed.  A possibility distinct from all of these symmetry-breaking phases is a Bose metal, i.e., a critical quantum liquid of bosons that does not break any symmetries yet has many gapless excitations.  Such Bose metal phases have been studied in detail previously, wherein the systems studied were compressible with respect to the addition of bosons~\cite{phillips,dbl_0,dbl_2leg}.

In this Letter, we will propose and study a unique phase of matter that has gapless collective excitations even though the phase is an incompressible insulator (Fig. \ref{fig:3ll}).  In particular, we will construct a trial wave function for this phase, suitable for use with variational Monte Carlo (VMC) calculations, and provide detailed comparisons to numerical results obtained using the density matrix renormalization group (DMRG) method to argue for the existence of such an exotic phase in a specific microscopic lattice model of bosons. We will focus our study here on the square lattice 3-leg ladder (see Fig.~\ref{fig:3ll}) at boson filling factor $\nu=1/3$ as a prototypical example of the physics of interest.  Even though our ultimate goal is to access novel two-dimensional (2D) phases, the more tractable ladder systems are very intriguing in their own right, and in our concluding remarks, we will propose clear extensions of our ideas that are aimed towards two dimensions.

\emph{Wave function}.---A wave function of bosons can be constructed by taking the product of two fermionic wave functions and evaluating them at the same coordinates (Gutzwiller projection):
\begin{equation}
\Psi_b(\mathbf{r}_1, ...,\mathbf{r}_M) = \Psi_{d_1}(\mathbf{r}_1, ...,\mathbf{r}_M)\,\Psi_{d_2}(\mathbf{r}_1, ...,\mathbf{r}_M),
\label{eqn:wf}
\end{equation}
which corresponds in terms of operators to a ``parton'' construction $b^\dagger=d_1^\dagger d_2^\dagger$ with the constraint $d^\dagger_1d_1=d^\dagger_2d_2=b^\dagger b$ \cite{dbl_0,dbl_2leg}.  In the case of interest here, where time reversal symmetry is present on a quasi-one-dimensional ladder, it is natural to put each fermion in a Slater determinant by occupying one-dimensional (1D) bands.  We stress that the $d_1$ and $d_2$ fermions need not have identical band filling configurations.

For the 3-leg ladder system (Fig.~\ref{fig:3ll}), there will be three bands of $d_1$ and $d_2$ fermions. At exactly $\nu=1/3$ filling, an interesting possibility arises that is special to this commensurate density: there are enough fermions to exactly fill one $d_2$ band completely, but still have up to three partially filled $d_1$ bands.  The resulting projected state of bosons is an exotic ``gapless Mott insulator'' (GMI):  the $d_2$ state is a band insulator while the $d_1$ state is a band metal. It is this phase that we will explore in detail in this Letter.

\emph{Model}.---The microscopic model for which we find the GMI phase consists of hard-core bosons hopping on a square lattice with a four-site ring-exchange interaction:
\begin{align}
\label{eqn:model}
H_{JK} = & -J\sum_{\rvec}(b^\dagger_{\rvec}b_{\rvec+\hat{x}}+\hc) -J_\perp\sum_{\rvec}(b^\dagger_{\rvec}b_{\rvec+\hat{y}}+\hc) \nonumber\\
&+\,K\sum_{\rvec}(b^\dagger_{\rvec}b_{\rvec+\hat{x}}b^\dagger_{\rvec+\hat{x}+\hat{y}}b_{\rvec+\hat{y}}+\hc),
\end{align}
where $b_{\rvec}$ destroys a boson at site $\rvec$. The $J$ and $J_\perp$ terms are the usual hopping terms for bosons in the directions parallel and perpendicular to the chains, respectively.   The ring term $K$ is the interaction that makes the phase diagram interesting. When $K>0$, which is the case studied here, the model has a sign problem thereby rendering quantum Monte Carlo calculations inapplicable.

The $K$ term causes a pair of bosons to hop without center-of-mass motion, as illustrated in Fig.~\ref{fig:3ll}. In this sense, it frustrates the attempt of the simple hopping term to cause the bosons to maximize their kinetic energy by condensing and yet at the same time disfavors the formation of a simple solid. A more compelling (albeit more abstract) motivation to study this model comes from a strong-coupling analysis of the U(1) gauge theory for the $d_1$ and $d_2$ fermions:  within this framework, one can reverse engineer Eq.~(\ref{eqn:model}) as a potential parent Hamiltonian for ground states of the form given by Eq.~(\ref{eqn:wf}) (see Ref.~\cite{dbl_0}).  Indeed, it was found that taking $d_1$ ($d_2$) to hop more easily along $x$ $(y)$ and then increasing this anisotropy effectively increases $K/J$.  At $\nu=1/3$ on the 3-leg ladder, one hence expects that as the $K$ term increases, the three $d_1$ bands should become almost degenerate, while the $d_2$ bands should become almost dispersionless and split apart.  This in turn naturally causes the $d_2$ fermions to go into a band insulating state and the $d_1$ fermions to go into a metallic state (see Fig.~\ref{fig:3ll}). This is exactly the situation discussed above in our description of the GMI phase, and is our primary motivation to search the phase diagram of the model, Eq.~(\ref{eqn:model}), for the GMI phase.

\emph{$3$-leg ladder}.---We henceforth consider Eq.~(\ref{eqn:model}) on the 3-leg ladder at filling factor $\nu=1/3$.  When $J \gg K,J_\perp$, the ground state is a superfluid (SF) with quasi-long-range order. On the other hand, when $J_\perp \gg K, J$, one boson on each rung goes into a zero $y$-momentum state, and the system is a conventional rung Mott insulator with a unique ground state and a gap to all excitations. The region where we expect the GMI phase is when the $K$ term dominates.

\begin{figure}[t]
\includegraphics[width=3.3in]{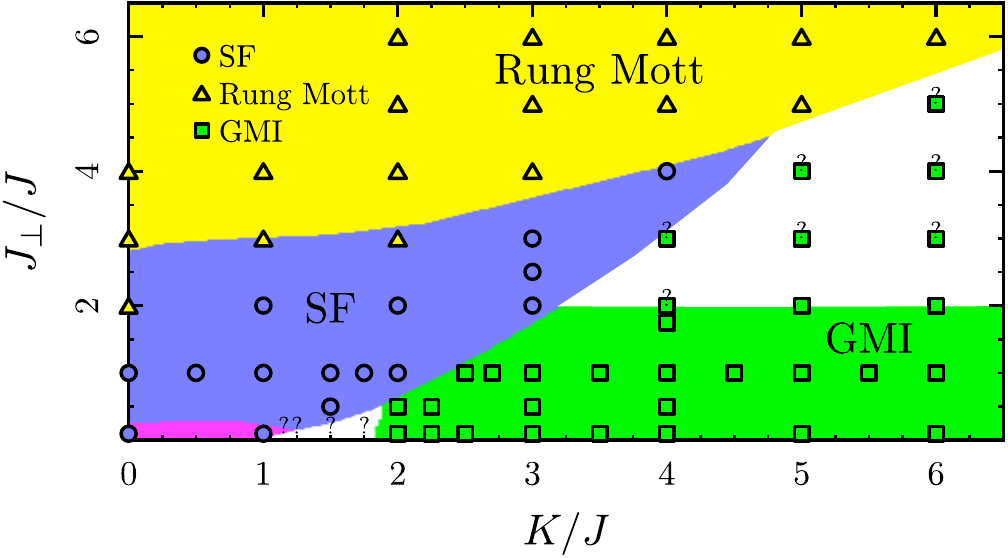}
\caption{(color online) Summary of our numerical results: phase diagram in the $K/J-J_\perp/J$ plane of the 3-leg ladder at $\nu=1/3$ filling as determined by DMRG and VMC calculations.
\label{fig:pd}}
\end{figure}

As we demonstrate below, we do in fact have extensive numerical evidence for the existence of an exotic GMI phase in the expected region of the phase diagram.  Specifically, in addition to the SF and fully gapped rung Mott state, we find a gapless Mott insulator with two 1D gapless modes.  Our results are summarized in Fig.~\ref{fig:pd}.  We first searched the phase diagram using the variational wave functions:  projected $d_1$ and $d_2$ bands with different band fillings (along with variational exponents on the two Slater determinants to give additional variational freedom \cite{dbl_2leg}) and Jastrow wave functions for the superfluid.  We then also scanned the phase diagram with DMRG at the points marked as circles, triangles, and squares in Fig.~\ref{fig:pd}, keeping between 2000 and 4000 states per block to ensure converged results.  Figure \ref{fig:pd} shows the phase diagram of $K/J$ vs. $J_\perp/J$ with colored regions (roughly) distinguishing the various phases.  These regions generally coincide with the VMC state that minimizes the energy at each point, although some adjustments have been made to incorporate strong conclusions from the DMRG.  The GMI phase has three partially occupied $d_1$ bands and one fully filled $d_2$ band---this is the phase of main interest.  The rung Mott phase is modeled well in the VMC calculations by a projected wave function in which both the $d_1$ and $d_2$ fermions fully fill the $k_y=0$ band.  There is also a SF phase and two unidentified regions (see below) marked as white.

\begin{figure}[t]
\vspace{-0.1in}
\subfigure{\includegraphics[width=3.5in]{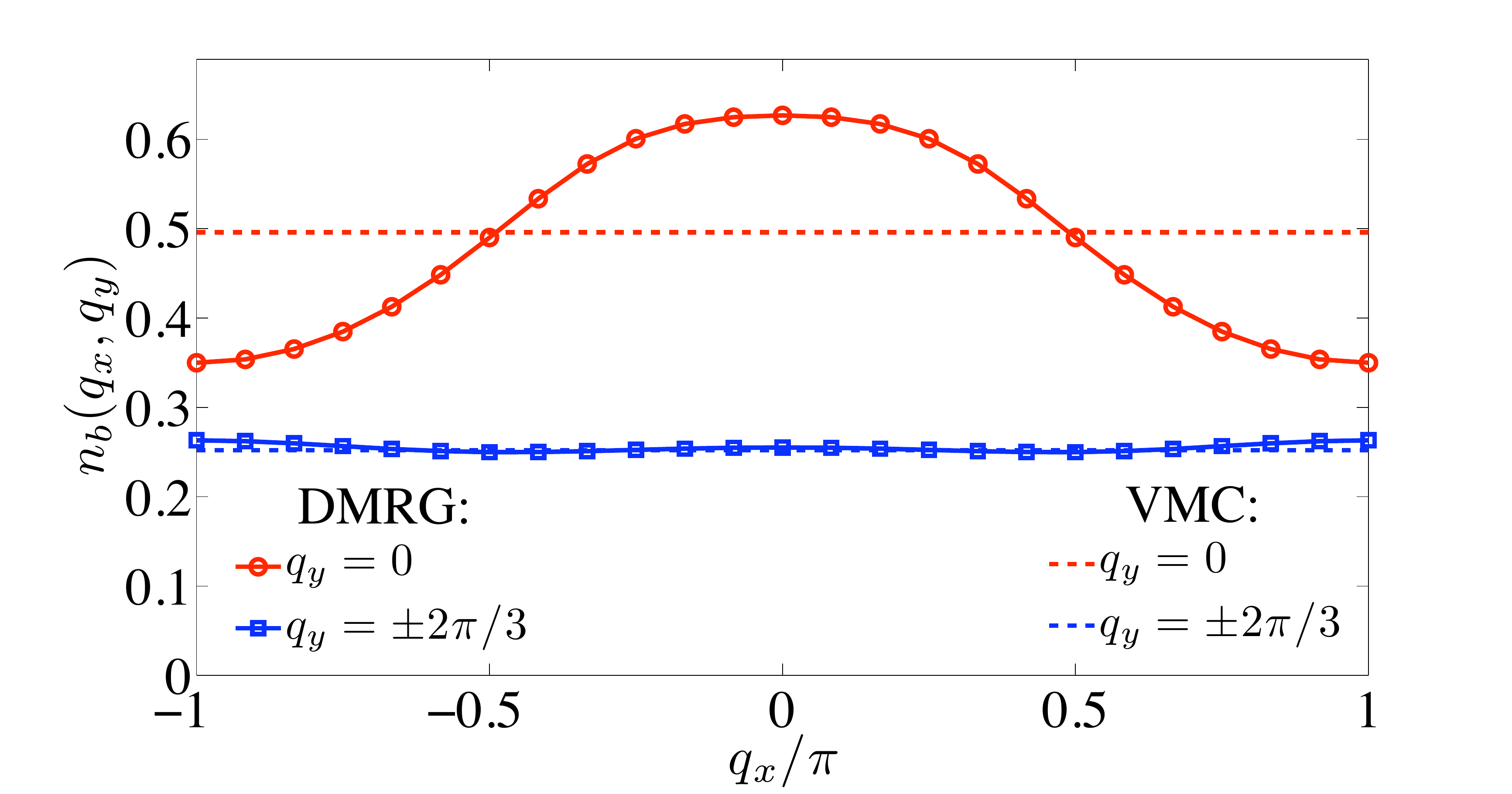}}\vspace{-0.1in}
\subfigure{\includegraphics[width=3.5in]{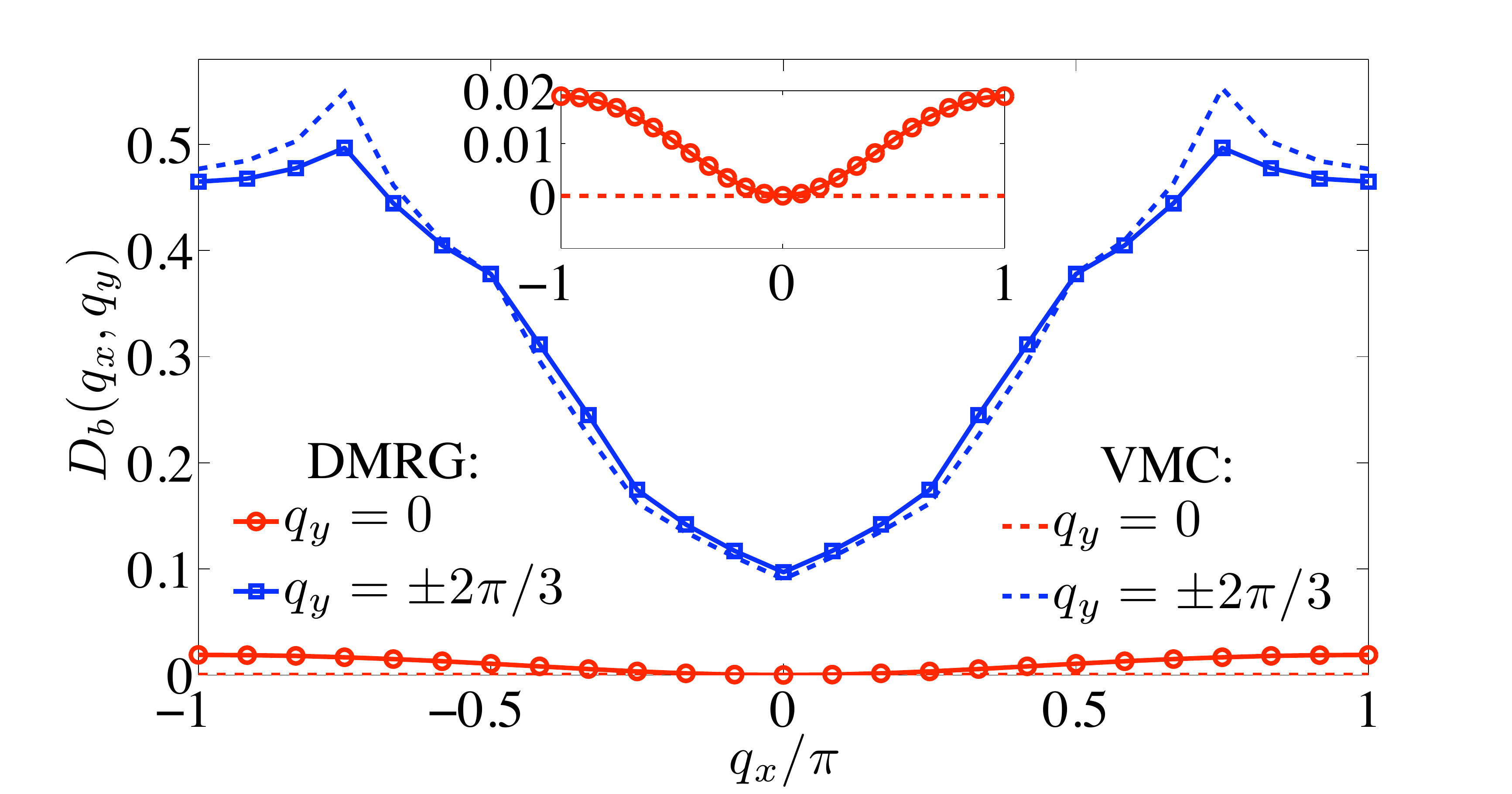}}\vspace{0.0in}
\subfigure{\includegraphics[width=3.2in]{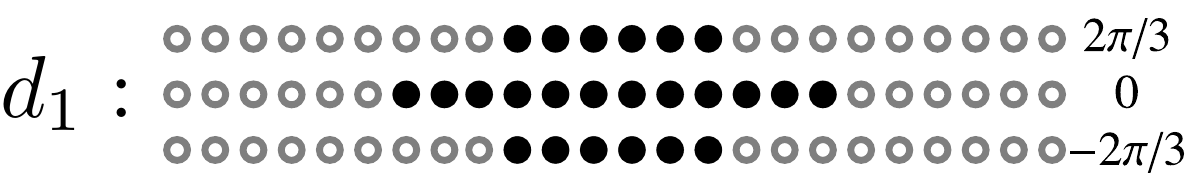}}
\caption{(color online) Comparison of DMRG and VMC calculations of the boson momentum distribution $n_b(\qvec)$ (top panel) and density-density structure factor $D_b(\qvec)$ (middle panel; inset is a zoom of $q_y=0$ curves) in the GMI phase of the 3-leg ladder at the point $K/J=2.7$ and $J_\perp/J=1.0$.  The lack of singular features in $n_b(\qvec)$ and $q_x^2$ dependence of $D_b(q_x,q_y=0)$ around $q_x=0$, both highlight the insulating nature of the GMI phase.  However, the density-density structure factor still ``sees'' the gaplessness of the phase due to the partially filled $d_1$ bands \cite{dbl_0}, hence the singular features in $D_b(q_x,q_y=\pm 2\pi/3)$.  These singularities can be identified with ``$2k_F$'' wave vectors in the $d_1$ fermion's band filling configuration (bottom panel) in which the $k_y=0$ band contains 12 fermions and the $k_y=\pm 2\pi/3$ bands each contain 6 fermions.  Of course, the $d_2$ fermion configuration (not shown) consists only of one fully filled $k_y=0$ band.
 \label{fig:dmrgvmc_cf}}
\end{figure}

We shall focus here on establishing the existence and properties of the GMI phase; these points are marked by squares (green) in Fig.~\ref{fig:pd}.  A comparison between VMC and DMRG calculations of the Fourier transformed single-boson Green's function and density-density correlation function, for a representative point in the GMI phase, is shown in Fig.~\ref{fig:dmrgvmc_cf} for a 3-leg system of length $L_x=24$.  Specifically, we plot the boson momentum distribution $n_b(\qvec)\equiv 1/(3 L_x) \sum_{\rvec,\rvec'}\langle b^\dagger_\rvec b_{\rvec'}\rangle e^{i\qvec\cdot(\rvec-\rvec')}$ and density-density structure factor $D_b(\qvec)\equiv 1/(3 L_x) \sum_{\rvec,\rvec'}\langle[ n_\rvec -\nu][n_{\rvec'} - \nu]\rangle e^{i\qvec\cdot(\rvec-\rvec')}$, where $n_\rvec\equiv b^\dagger_\rvec b_\rvec$.

In the proposed GMI phase, since the $d_2$ fermion is gapped, there is a gap to any excitation that carries nonzero boson number.  This implies a single-boson Green's function that decays exponentially in real space and thus a boson momentum distribution with no singular features; the latter is clearly demonstrated in the top panel of Fig. \ref{fig:dmrgvmc_cf} in both the DMRG and VMC.  Because of the fully filled $d_2$ band, the VMC state contains exactly one boson on each rung and hence gives completely flat $n_b(q_x, q_y=0,\pm 2\pi/3)$, while the true ground state (as obtained by DMRG) does contain some short-distance fluctuations.  Within the GMI, the $\qvec=0$ mode occupation, $n_b(\qvec=0)$, does not increase with increasing system size in the DMRG, and a finite-size study of the one- and two-boson gaps indicates that these gaps attain rather large, finite values in the $L_x\rightarrow\infty$ limit \cite{epaps}.  All of these findings are consistent with the insulating properties of the GMI phase.

Measurement of the density-density correlation function gives compelling evidence for the presence of the GMI phase, in particular, its gapless nature and incompressibility.  In the VMC wave function, $D_b(\qvec)$ contains contributions from gapless particle-hole excitations arising from partially filled bands of either fermion species ($d_1$ only in the case of the GMI) and hence has singular structure at various momenta \cite{dbl_0}.  Indeed, the DMRG calculation of the density-density structure factor displays singular features that are impressively well reproduced by the corresponding VMC calculation in which the lowest energy state has a band filling configuration shown in the bottom panel of Fig.~\ref{fig:dmrgvmc_cf}.  In this case, the main singularities in $D_b(\qvec)$ at $(q_x,q_y)=(\pm 3\pi/4,\pm 2\pi/3)$ can be identified with the ``$2k_F$'' wave vectors associated with moving a $d_1$ fermion from the $k_y=0$ band to the $k_y=\pm 2\pi/3$ bands (going across the bands); the other singularities can similarly be identified with other ``$2k_F$'' moves that are allowed.  In particular, singularities at $q_y=0$ are suppressed by $d_2$:  the VMC wave function has exactly one boson on each rung which produces $D_b(q_x,q_y=0)=0$.  The DMRG calculation, on the other hand, gives small but finite $D_b(q_x,q_y=0)$ with $q_x^2$ dependence around $q_x=0$.  This result directly implies that the system is an incompressible insulator and rules out the possibility of a superfluid, either conventional or paired.  These results are general for the entire GMI phase in Fig. \ref{fig:pd}, except that the locations of the singularities in $D_b(q_x,q_y=\pm 2\pi/3)$ can be changed by tuning $K/J$ and/or $J_\perp/J$ to alter the resulting $d_1$ band filling configuration.

\begin{figure}[t]
\vspace{-0.1in}
\includegraphics[width=3.65in]{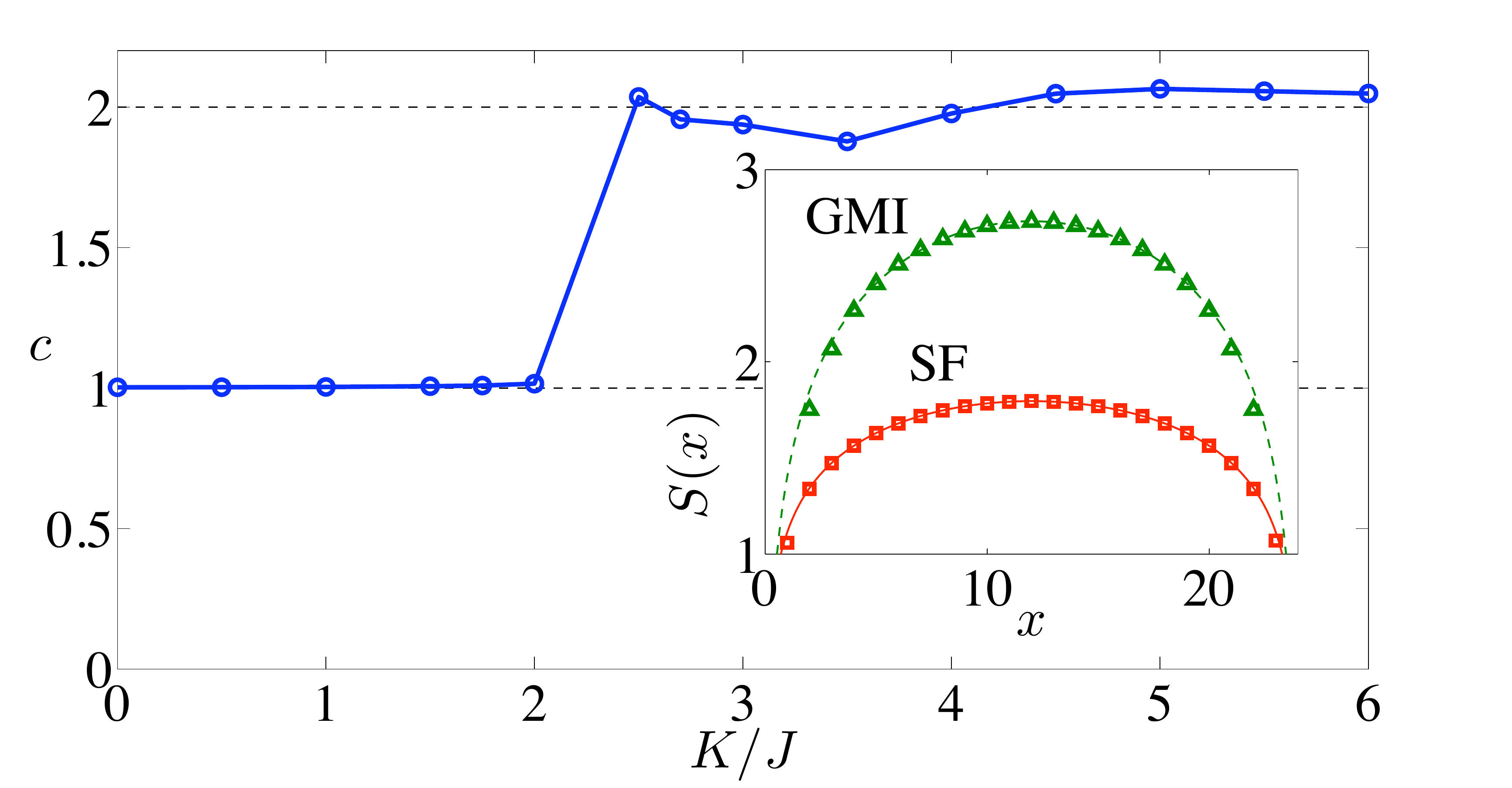}
\caption{(color online) Effective central charge $c$ as inferred from scaling of the entanglement entropy versus subsystem size $S(x)$.  For the 3-leg ladder, we fit the von Neumann entanglement entropy of leftmost contiguous blocks of size $3x$ to the scaling form $S(x)=(c/3)\ln\left[(L_x/\pi)\sin\left(\pi x/L_x\right)\right]+A$ to extract $c$ \cite{cardy}.  In the main plot, $c$ is plotted versus $K/J$ at $J_\perp/J=1.0$ for a system of length $L_x=24$; the jump from $c\simeq 1$ to $c\simeq 2$ at the SF-GMI boundary is striking.  In the inset, we plot $S(x)$ in both the SF ($K/J=1.0$) and GMI ($K/J=2.7$) phases; the points are the DMRG data, while the curves are the fits to the scaling form.  \label{fig:ee}}
\end{figure}

A crucial aspect of the GMI phase is the expectation that it has two 1D gapless modes at low energy: at the mean-field level, Fig.~\ref{fig:3ll}, there are naively three 1D gapless modes corresponding to the three partially filled $d_1$ bands.  We must, however, enforce a constraint such that $d_1^{\dagger}(\rvec)d_1(\rvec)=d_2^{\dagger}(\rvec)d_2(\rvec)$ to realize the physical bosons.  This is accomplished by means of a gauge field along with corresponding (opposite) gauge charges attached to each flavor of parton.  In the strong-coupling limit, one linear combination of the original gapless modes is rendered massive leaving only two gapless modes~\cite{dbl_2leg,followup}.  We have calculated the entanglement entropy of the ground state wave function with DMRG and extracted from it the effective central charge \cite{cardy,sbm_2leg}, as shown in Fig.~\ref{fig:ee}. In the superfluid phase, we find that the central charge $c\simeq 1$ as expected for a single gapless boson density mode in this phase. As one approaches the transition into the GMI phase, one finds a sharp increase to $c\simeq 2$, providing strong evidence that the GMI phase has two 1D gapless modes.  That is, as elucidated by Fig.~\ref{fig:ee}, even though the GMI is an insulator, it actually contains more entanglement and more 1D gapless modes than the SF.

In addition to the phases mentioned above (SF, rung Mott, and exotic GMI), there are two unidentified (white) regions in the phase diagram of Fig.~\ref{fig:pd}:  a small region at intermediate $K/J$ and very small $J_\perp/J$ and a larger region above the GMI.  The corresponding DMRG points are marked as question marks and squares (green) with a question mark, respectively.  In the latter region, VMC favors a GMI ground state, but the features of the GMI in the DMRG fail to persist, e.g., the singularities in $D_b(q_x,q_y=\pm 2\pi/3)$ smoothen and the calculated central charge drops below $c\simeq2$.  An analysis of the spectral gap versus $L_x$ using exact diagonalization and DMRG reveals that the system is either gapless or has a very small gap in this region \cite{epaps}, without any obvious ordering.

\emph{Extensions}.---A natural extension of our findings is to the general $N$-leg ladder at $\nu=m/N$, with integer $m$.  In such systems, a GMI phase analogous to the one discussed above is a clear ground state candidate.  We leave for future work a numerical investigation using DMRG to assess the stability of GMI phases in these systems; although we can report that on the 2-leg ladder at $\nu=1/2$ the GMI is likely unstable in our model \cite{followup}, a result which partially motivated the present study.

Another interesting extension of our results is to structures that consist of $N$ layers of 2D systems at a general filling of $\nu=m/N$, e.g., a three-layer system at $\nu=1/3$.  In such a system, the $d_2$ fermions can exactly fill $m$ of the $N$ 2D bands resulting in a gapless Mott insulating state.  Theoretically, this state is stable and can exist in two dimensions. We expect that an appropriate model that realizes this physics would have substantial ring exchange on the plaquettes between the layers, as this interaction will give the necessary disparity in hopping to keep the $d_1$ system metallic while forcing the $d_2$ system into a band insulating state. Thus, we are hopeful that an appropriate $JK$ model will realize this quasi-2D analogue.

This work was supported by the NSF under Grants No. DMR-0529399 (M.S.B., R.V.M., and M.P.A.F.) and No. DMR-0907145 (O.I.M.); the DOE under Grant No. DE-FG02-06ER46305 (D.N.S.); and Microsoft Station Q  (R.V.M. and R.K.K.).

\end{document}